\documentclass[conference]{IEEEtran}
\IEEEoverridecommandlockouts
\usepackage{cite}
\usepackage{amsmath,amssymb,amsfonts}
\usepackage{algorithmic}
\usepackage{graphicx}
\usepackage{textcomp}
\usepackage{xcolor}
\usepackage[caption=false,font=footnotesize]{subfig}
\usepackage{comment}
\usepackage{booktabs}
\def\BibTeX{{\rm B\kern-.05em{\sc i\kern-.025em b}\kern-.08em
    T\kern-.1667em\lower.7ex\hbox{E}\kern-.125emX}}

\title{Cascading Contingency Analysis of Modern Power Systems Using a Stochastic Hybrid Systems Framework

\thanks{This work is supported by Wayne State University Mobility Accelerator Innovation Network Grant
through the Office of Commercialization. \\
* Corresponding Author}
}

\author{\IEEEauthorblockN{Masoud H. Nazari*,~\IEEEmembership{Senior Member,~IEEE}}
\IEEEauthorblockA{
\textit{Electrical and Computer Engineering} \\
\textit{Wayne State University} \\ 
\textit{Detroit, USA}\\
\textit{Electrical and Computer Engineering} \\
\textit{University of Waterloo, Waterloo, Canada} \\
masoud.nazari@wayne.edu}
\and
\IEEEauthorblockN{Hamid Varmazyari,~\IEEEmembership{Student Member,~IEEE}}
\IEEEauthorblockA{
\textit{Electrical and Computer Engineering} \\
\textit{Wayne State University} \\ Detroit, USA\\
varmazyari.h@wayne.edu}
}
\begin{document}
\maketitle

\begin{abstract}
This paper presents a cascading contingency analysis and risk assessment framework that incorporates probabilistic mode transitions motivated by stochastic hybrid systems (SHS) concepts together with power-flow-based stress evaluation. A two-stage modeling approach is adopted in which sensitivity-based analysis is first used to identify candidate cascading transitions, followed by DC power flow re-evaluation to accurately capture post-contingency stress redistribution, islanding behavior, and load shedding. A probabilistic propagation model is constructed using mode-dependent transition probabilities derived from line loading conditions. A Cascading Risk Index (CRI) is introduced to quantify the expected load shedding accumulated over cascade progression. Case studies on the IEEE 39-bus system demonstrate that the proposed framework captures diverse cascading routes under multiple initiating outages and identifies critical propagation patterns that are not revealed by N–1/N–2 screening. The results highlight the importance of SHS modeling in assessing cascading risk and providing a scalable framework for resilience evaluation in modern power systems.
\end{abstract}

\section{Introduction}
\label{Sec1}

\IEEEPARstart{C}{ascading} failures are complex events in which the outage of one or more components propagates through the power grid, leading to widespread disruptions \cite{Cascad16, ZAKARIYA2023108928}.
Notable examples include the April 2025 Spain-Portugal blackout, where system oscillations and a set of rapid sequential disconnections resulted in 15 GW of generation loss. Large disconnections from solar installations in southern Spain contributed to destabilizing the grid, which already had low system inertia and weak interconnection with France, affecting millions of people in the Iberian Peninsula \cite{SPPT1}. %Similarly, the North American Electric Reliability Corporation (NERC) reports have documented multiple inverter-based resources (IBR)-related failures, such as the 2021 Texas blackout and 2016 Blue Cut Fire event, where inadequate IBR fault ride-through capabilities and insufficient grid-forming controls exacerbated outages \cite{SPPT2}.

Modern power systems are increasingly subject to dynamic and stochastic dependencies among components, as well as emergent behaviors induced by high inverter-based resources (IBR) penetration. Conventional reliability standards such as the $N\!-\!1$ and $N\!-\!2$ contingency criteria cannot adequately capture dynamic propagation of failures and multi-component dependencies. Modern grids require analytical frameworks that account for:
\begin{enumerate}
    \item Dynamic and stochastic dependencies among components that can result in cascading events,
    \item Physical mechanisms such as thermal overloads, protection miscoordination, and stress redistribution, and
    \item Emergent behaviors induced by high distributed energy resources (DER) and IBR penetration.
\end{enumerate}

We develop an SHS-based cascading contingency framework that captures both discrete contingency transitions and continuous power flow dynamics under uncertainty within a linear, mode-dependent dynamical representation. The model builds on our earlier work in \cite{cd2, 11272230}.

To analyze the propagation of contingencies, we propose a probabilistic cascading contingency analysis and risk assessment framework that focuses on the post-contingency evolution of failures across the network. This constitutes the main contribution of the paper, as it explicitly captures system dynamics and the propagation of stress through the network.

%We develop an SHS-based cascading contingency framework capable of modeling both discrete contingency transitions and continuous power flow dynamics under uncertainty within a linear, per-mode dynamical representation. The model is based on our earlier works in \cite{11225525, 11272230}. To detect subsequent propagation of contingencies, we develop a probabilistic cascading contingency analysis and risk assessment framework that focuses on the post-contingency evolution of failures across the network. This forms the main contribution of the paper, as it explicitly captures the physical dynamics and stress propagation.  

Unlike traditional contingency assessment methods that treat failures as isolated or independent events \cite{mishra2012contingency, TraCascd1, TraCascd2}, the proposed framework models how power system stressors---such as thermal overloads, and protection miscoordination---influence the progression of cascading outages. By embedding these interactions in an SHS formulation with linearized dynamics, the model can reveal hidden dependencies among components, quantify mode-dependent transition probabilities, and identify early-stage indicators of cascading chain reactions.

Nonlinear phenomena (e.g., voltage collapse dynamics and nonlinear protection characteristics) are acknowledged but not explicitly modeled in this work. The present formulation adopts DC power flow representations within each contingency mode. Incorporating nonlinear models, such as AC power flow, is reserved for future research.

%Nonlinear phenomena (e.g., full voltage collapse dynamics, nonlinear protection characteristics) are acknowledged but not modeled in this paper; the present development assumes DC power flow dynamics per contingency mode. Nonlinear models, such as AC power flow will be subject of the future work.  

The remainder of this paper is organized as follows. Section \ref{Sec2} introduces the modeling framework for power systems and the contingency propagation model. Section \ref{Sec3} presents contingency risk assessment via Markov chain (MC) and graph-based methods. Section \ref{Sec4} gives simulation studies on the IEEE 39-bus case, and Section \ref{Sec5} concludes the paper by summarizing the main findings and outlining future research directions.  

\section{Modeling Framework}
\label{Sec2}
\subsection{Stochastic Hybrid System Representation} 
%and Markov Modeling
Within each contingency mode, the power system dynamics are taken to be adequately represented by a linear, time-invariant (LTI) model. Let the power system be represented by a continuous state vector $x(t)$ and a discrete mode $\alpha_k \in \ \mathcal{M} = \{1,2,\ldots,M\}$ denoting the active contingency configuration at time $t_k$. A general SHS formulation can be used to represent such interactions \cite{cd2}:
\begin{align}
    \dot{x}(t) &= A^{\alpha_k} x(t) + B^{\alpha_k} u(t) + w(t), \\
    y(t) &= C^{\alpha_k} x(t) + v(t),
\end{align}
where $w(t)$ and $v(t)$ are stochastic disturbances, and each mode $\alpha_k$ corresponds to a specific set of failed components (lines, transformers, or generators).
To make the framework analytically and computationally tractable, we adopt the following distributional assumptions: $w(t)$ and $v(t)$ are modeled as zero-mean Gaussian white noise processes with known (or estimated) covariance matrices $Q$ and $R$ respectively. Under this assumption, the state $x(t)$ remains Gaussian within each mode. 
%permitting closed-form propagation and consistent downstream probabilistic analysis. 
Extensions to bounded or heavy-tailed disturbances are left for future work.

 This work adopts a reduced-order representation in which the continuous dynamics are not explicitly simulated. DC power flow analysis is used to characterize system conditions within each contingency mode and capture active power redistribution and thermal loading. %which are the dominant drivers of cascading propagation in transmission systems.

The mode transitions are modeled using an MC characterized by transition probabilities $P_{ij}$:
\begin{equation}
P_{ij} = \Pr(\alpha_{k+1}=j \mid \alpha_k=i),
\label{MCP}
\end{equation}
which captures the likelihood of subsequent outages given the current contingency. %{\color{red}This formulation assumes that the current contingency mode, together with the associated power flow conditions (e.g., line loadings), sufficiently represents the system state. Under this assumption, the probability of subsequent outages depends only on the current mode, enabling tractable probabilistic modeling of cascading sequences.}
The probability of each MC transition following a contingency is computed using power flow sensitivity analysis to assess whether thermal limits and other power network constraints are violated after each state transition. This approach captures key physical realities—such as thermal inertia and protection timing—that introduce history dependence into the cascading process. 

The transition probabilities are derived using a two-stage approach. Power flow sensitivity analysis is first used to identify candidate outage lines based on their impact on network stress. DC power flow evaluation is then performed to compute post-contingency line loadings, from which thermal limit violations and overload margins are used to quantify the likelihood of subsequent outages.
Thus, we can assume that the current contingency mode, together with the associated power flow conditions (e.g., line loadings), provides sufficient information to determine the next-step transition probabilities.

These transition probabilities encode physical mechanisms---such as thermal overloads, protection misoperations, and stress redistribution. The Markovian switching process characterizes the probabilistic propagation of cascading outages by capturing how an initial disturbance can trigger progressive failures through physical interactions. Finally, the probability of a cascading route $\mathcal{C} = (i_0, i_1, \dots, i_K)$ is computed as the product of transition probabilities along the sequence:

\begin{equation}
\mathbb{P}(\mathcal{C}) = \prod_{t=0}^{K-1} P_{i_t i_{t+1}}.
\end{equation}

The resulting SHS cascading framework enables systematic modeling of cascading sequences while maintaining computational tractability for large-scale contingency analysis.

\subsection{Load Shedding and Operating Constraints}

Post-contingency system conditions are evaluated using a slack-limited DC power flow model. Let $P_{\text{load}}$ denote the total system demand and $P_{\text{gen}}$ denote the available generation, including a slack bus with maximum capacity $P_{\text{slack}}^{\max}$. When the total demand exceeds available generation, load is proportionally reduced to maintain power balance. The served load fraction is defined as

\begin{equation}
\lambda = \min\left(1, \frac{P_{\text{gen}}}{P_{\text{load}}}\right),
\end{equation}
and the served load is given by $P_{\text{served}} = \lambda P_{\text{load}}$, with corresponding load shedding $P_{\text{shed}} = (1-\lambda)P_{\text{load}}$.

To evaluate the impact of network constraints, DC power flow is used to compute line flows under each contingency mode. Let $L_\ell$ denote the real power flow on line $\ell$ under the current contingency mode (i.e., post-contingency), as obtained from DC power flow, and $\bar{L}_\ell$ denote the corresponding thermal rating of that line. The loading ratio $\frac{L_\ell}{\bar{L}_\ell}$ quantifies the proximity of each line to its thermal limit.

Based on these line loadings, a system-level stress metric is defined as

\begin{equation}
S = \max_{\ell \in \mathcal{L}} \left( \frac{|L_\ell|}{\bar{L}_\ell} \right),
\end{equation}
which represents the most critical loading condition in the network. This stress metric captures the extent to which the system operates under constrained or overloaded conditions following a contingency.

Let $S_{\text{safe}}$ and $S_{\text{crit}}$ denote the safe and critical
loading thresholds. The candidate
served-load fraction at cascade step $k+1$ is computed as

\begin{equation}
\lambda_{k+1}^{\text{candidate}}
=
\lambda_k
\left[
1-\beta
\frac{S-S_{\text{safe}}}
{S_{\text{crit}}-S_{\text{safe}}}
\right],
\qquad
S_{\text{safe}} < S < S_{\text{crit}},
\end{equation}

where $\beta \in [0,1]$ is a degradation coefficient that
controls the severity of load reduction under stressed
operating conditions. Larger values of $S$ therefore lead
to lower candidate served-load fractions. 
%reflecting the increased vulnerability of heavily loaded networks to cascading effects.

In addition to generation adequacy, this stress metric is used to further adjust the served load. 
%reflecting the fact that highly stressed systems are more vulnerable to protection actions and operational limitations. 
As $S$ increases beyond safe operating levels, the served load is progressively degraded to account for the increased likelihood of cascading effects.

To reflect the irreversible nature of cascading failures, the served load is constrained to be non-increasing along each cascade path. If $\lambda_k$ denotes the served load fraction at cascade step $k$, then

\begin{equation}
\lambda_{k+1} = \min\left(\lambda_k, \lambda_{k+1}^{\text{candidate}}\right).
\end{equation}

Finally, when network islanding occurs, each island is evaluated independently. Only the island containing the slack bus is assumed to remain energized, while loads in islands without generation are considered fully shed. This ensures that the served load reflects both generation availability and network connectivity constraints.

\subsection{Contingency Propagation Model}

The contingency propagation model determines which additional line outages may occur after a given contingency mode and assigns probabilities to those candidate transitions. At each cascade step, candidate subsequent outages are restricted to in-service lines whose loading exceeds their thermal rating. Therefore, for the current contingency mode, a line $\ell$ is considered eligible for propagation if \begin{equation}
\frac{|L_\ell|}{\bar{L}_\ell} > 1,
\end{equation}
To capture the relationship between component stress and the likelihood of failure, we model
the propagation probability as
\begin{equation}
p_{\text{prop}}^{(\ell)} =
\sigma\!\left(\frac{|L_\ell|-\bar{L}_\ell}{\Delta L}\right),
\label{Prop}
\end{equation}
where $\sigma(\cdot)$ is the logistic (sigmoid) activation function and $\Delta L$ is a sensitivity
parameter that determines how sharply the probability increases as loading approaches the thermal limit.
This formulation reflects the physical intuition that lightly loaded components have negligible probability of failure, whereas overloaded components are increasingly likely to trip. The sigmoid mapping provides a smooth transition between safe and highly stressed operating conditions, while capturing uncertainty in outage behavior. Moreover, uncertainty in system behavior is reflected through the probabilistic mapping between line loading and outage likelihood, which captures variability in operating conditions, protection responses, and model approximations.

To integrate the propagation model with the SHS, we adopt the following. For a mode $i$ and a candidate next mode $j$ that differs from $i$ by the outage of a single line $\ell$, we define the transition weight $w_{ij}$ based on the component-level propagation probability for line $\ell$ evaluated at the current loading.  Specifically, the propagation probability $p_{\text{prop}}^{(\ell)}$ is a function of the current loading condition, and the transition weights are defined as\begin{equation}
w_{ij} =
\begin{cases}
p_{\text{prop}}^{(\ell)}, & \text{if line } \ell \text{ is eligible for propagation},\\
0, & \text{otherwise}.
\end{cases}
\end{equation}where $\ell$ is the line whose outage differentiates mode $j$ from mode $i$.

If we have transition weights $w_{ij} \ge 0$ obtained from loading-dependent propagation probabilities, the normalization to enforce a Markov transition matrix is

\begin{equation}
P_{ij} = \frac{w_{ij}}{\sum_{k \in \mathcal{M}} w_{ik}}
\quad \Rightarrow \quad
\sum_{j \in \mathcal{M}} P_{ij} = 1, \quad \forall i,
\end{equation}

provided that
\begin{equation}
\sum_{k \in \mathcal{M}} w_{ik} > 0.
\end{equation}
If no line satisfies the overload condition, then no further transitions are possible, and the current mode is treated as an absorbing state.
In the baseline formulation, each Markov transition corresponds to the outage of at most one additional line. Hence, simultaneous multi-line failures within a single time step are not modeled explicitly and are approximated as a sequence of single-line outages occurring over successive transitions. This approximation is reasonable when the Markov time step is sufficiently small relative to the cascade propagation time scale. If simultaneous outages must be represented explicitly, the mode space $\mathcal{M}$ can be expanded to include joint-outage states and the transition weights can be defined using a joint probability model over subsets of lines. Such an extension, however, leads to a combinatorial increase in the state space and is therefore not considered in this paper.

We directly obtain the line loading from power flow evaluation within each contingency mode. Consequently, the component-level propagation probability is determined as a function of the current loading condition.
In \eqref{Prop}, the parameter $\Delta L$ controls the sharpness of the failure transition. When historical outage or outage management system (OMS) data are available, $\Delta L$ can be estimated via
logistic regression of next-step tripping events $Y$ on the overload margin
\[
m = |L_\ell| - \bar{L}_\ell,
\]
yielding
\[
\Delta L = \frac{1}{\hat{\beta}_1},
\]
where $\hat{\beta}_1$ is the estimated slope coefficient (with an optional
intercept term to capture baseline tripping risk).
This probabilistic representation provides a flexible and physically interpretable mechanism for
quantifying cascading risk. 
%within the proposed cascading framework.

\section{Contingency Risk Assessment}
\label{Sec3}
\subsection{Markov Chain}

To evaluate cascading outage risk under uncertainty, we employ an MC framework that
represents the probabilistic propagation of component failures \cite{MC16}. %Let the system state at step $k$ be denoted by a discrete contingency mode $\alpha_k \in \{1,2,\ldots,M\}$, where each mode corresponds to a specific configuration of failed and operational components. 
The MC transition matrix $P=[P_{ij}]$ encapsulates the likelihood of all possible
multi-stage failure trajectories.

During each simulation, the chain evolves according to $\alpha_{k+1} \sim P(\cdot \mid \alpha_k),$
and transitions continue until the system reaches an absorbing state $\alpha^\star$, representing a mode
in which no additional outages occur (i.e., $\alpha_{k+1}=\alpha_k$). 
%By generating a large number of sampled trajectories, the framework estimates both the probability distribution of cascading depths and the expected severity of multi-stage contingency sequences. This provides an empirical measure of risk that naturally incorporates uncertainties and physical dependencies.
This MC representation enables the evaluation of cascade progression probabilities and the identification of high-risk contingency sequences. It also captures both physical stress propagation and uncertainty in outage occurrence.

\subsection{Graph-Based Methods}
Graph-theoretic vulnerability metrics complement the MC–based probabilistic framework by quantifying the structural importance of transmission lines and buses. \textit{Betweenness centrality} measures how often a line lies on shortest electrical paths, identifying critical flow corridors \cite{Btw1}; \textit{electrical distance} metrics, based on weighted network impedances, characterize how tightly coupled different regions of the grid are and highlight subnetworks where local disturbances can propagate rapidly.

In a DC power flow setting, the failure of one line redistributes flow on other lines via Line Outage Distribution Factors (LODFs). To avoid a fully dense propagation matrix, we define the graph-based contingency propagation matrix $G$ using a quantitative threshold $\epsilon > 0$:

The matrix $G$ encodes topological and flow-coupling dependencies among component failures and is defined as follows:
\begin{equation}
G_{ij} =
\begin{cases}
1, & \text{if the failure of line } i \text{ increases the stress} \\[4pt]
 & \text{or affects line } j \text{ (e.g., } |\mathrm{LODF}_{j,i}| > \epsilon),\\[4pt]
0, & \text{otherwise}.
\end{cases}
\end{equation}
The threshold $\epsilon$ is chosen so that only non-negligible stress transfers are counted. The matrix $G$ is used as a structural constraint on the propagation model: only line outages that satisfy $G_{ij}=1$ are considered feasible transitions.
The resulting cascade pathways and risk metrics may depend on both the threshold $\epsilon$ and the underlying LODF model. 
%used to compute stress-transfer relationships. 
In particular, smaller values of $\epsilon$ produce a denser graph $G$, allowing more candidate propagation paths and typically yielding higher estimated cascade risk, whereas larger values of $\epsilon$ produce a sparser graph and a more selective propagation structure. 
%
%The sensitivity of the risk metrics (e.g., cascade size, expected load shed, or CRI) to $\epsilon$ should be reported by evaluating the model over a range of thresholds around the nominal value. Likewise, since the entries of $G$ depend on the adopted LODF approximation, the resulting risk estimates should be compared under alternative LODF formulations, such as the standard DC load-flow-based LODF model versus other linearized or higher-fidelity approximations when available. This sensitivity analysis helps quantify the robustness of the inferred cascade structure and ensures that the reported risk metrics are not artifacts of a particular masking threshold or flow model.}

Matrix $G$ constrains the MC transitions by limiting feasible next-state modes to those that respect electrical connectivity, proximity, and load-transfer relationships, so that the support of $P_{ij}$ is restricted to physically meaningful outage propagation paths.
 The sensitivity of the risk metric to the LODF model is defined as:
\begin{equation}
S_{\mathcal{L}}^{R}(\mathcal{L}_{a})
=
\frac{R(\epsilon,\mathcal{L}_{a})-R(\epsilon,\mathcal{L}_{\mathrm{DC}})}
{R(\epsilon,\mathcal{L}_{\mathrm{DC}})}.
\end{equation}

Here, $R(\epsilon,\mathcal{L})$ denotes the risk metric computed using the SHS framework under contingency level $\epsilon$ and LODF formulation $\mathcal{L}$. The term $\mathcal{L}_{\mathrm{DC}}$ represents the reference DC load-flow-based LODF model, while $\mathcal{L}_{a}$ denotes an alternative LODF approximation.
This sensitivity measure quantifies the relative deviation of the risk metric from the DC-based benchmark and enables the identification of transmission paths that are more susceptible to cascading propagation \cite{MN-epsr26}.

%\sout{Combining MC sampling with graph-based structural indicators yields a comprehensive assessment of cascading vulnerability:} {\color{red}Combining MC framework with graph-based structural indicators yields a comprehensive assessment of cascading vulnerability:} \sout{the chain captures stochastic failure propagation from thermal overloads and protection dynamics,} {\color{red}the Markov model captures probabilistic failure propagation driven by overload conditions,} while the graph-theoretic analysis reveals network features that enable or constrain propagation. Together, they quantify both probabilistic risk and structural drivers of cascading outages \cite{MN-epsr26}.

\subsection {Cascading Risk Index}
To quantify the likelihood and severity of cascading failures, we define a
CRI that incorporates both the probabilistic evolution of outage sequences and their resulting system
impact \cite{CRI}. 
The absorption probabilities into
terminal (no-more-outages) states provide the overall likelihood that a specific cascade sequence
occurs. 
Using these mode-dependent probabilities, the CRI is defined as
\begin{equation}
    CRI = \sum_{i=1}^{M} p_i \cdot L_i,
\end{equation}
where $p_i$ is the probability that the system reaches contingency state $i$ under the MC, and
$L_i$ is the associated severity metric. The impact $L_i$ may represent load shed, voltage collapse extent, generator tripping, frequency deviation, or topological fragmentation. 

In this paper, a single, consistent severity metric is adopted for all states: $L_i$ denotes the total load shed (MW) associated with contingency state $i$. Other choices have different units and physical meaning; a multi-objective or normalized aggregation could be used in future work.
In addition, the CRI is an expected-value metric, i.e., it represents the expected load shed across contingency states. More risk-averse measures, such as Conditional Value-at-Risk (CVaR), may be considered to capture tail-risk behavior in future work.

The CRI incorporates stochastic transition probabilities derived from the SHS-based framework, correlated outage behaviors arising from load redistribution and protection-system interactions, and multi-stage failure propagation pathways encoded in $P$. By capturing both the evolving system state and the interdependent mechanisms governing cascading outages, the CRI provides a comprehensive assessment of system resilience under dynamic disturbances. This enables operators to evaluate not only short-term operational stability but also long-term system vulnerability as disturbances propagate through the network.
In contrast, traditional $N$--$k$ security criteria treat outages as independent events and neglect the dynamic, mode-dependent nature of cascading failure propagation \cite{N-k1}. Consequently, the proposed CRI offers a more realistic and rigorous metric for assessing power system resilience.

%The CRI integrates stochastic transition probabilities from the SHS-based modeling, correlated outage behaviors due to load redistribution and protection coupling, and multi-stage failure propagation pathways encoded in $P$. 
%It captures both the evolving system state and the interdependent mechanisms that drive cascading failures, and enables operators to evaluate short-term stability and long-term vulnerability under dynamic disturbances.
%In contrast, $N$--$k$ security criteria assume independent outages and ignore the dynamic, mode-dependent nature of failure propagation \cite{N-k1}. Thus, the CRI proposed in this paper offers a more realistic and rigorous measure of resilience. 

\section{Simulation Results}
\label{Sec4}

\begin{figure}
    \centering
    \includegraphics[width=0.9\linewidth]{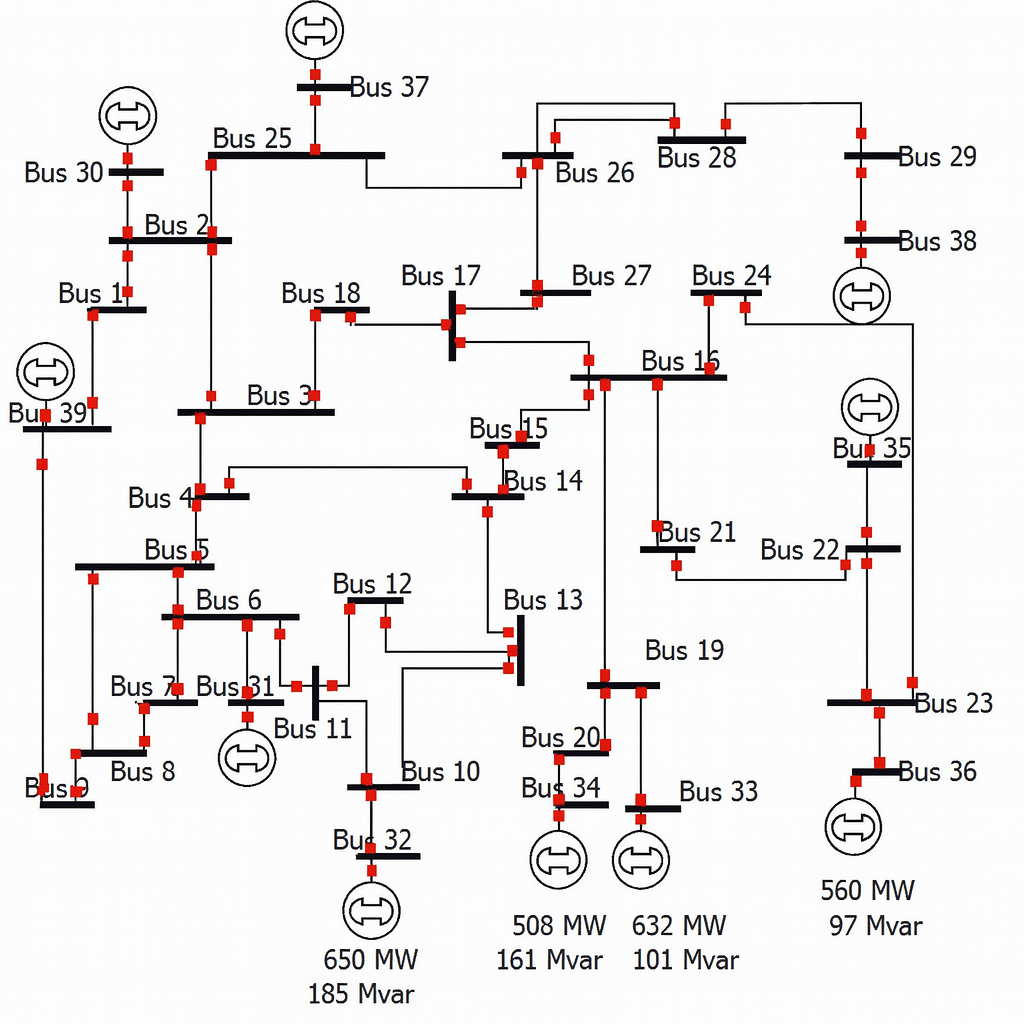}
    \caption{One-line diagram of the IEEE 39-bus system \cite{pai89}.}
    \label{fig:39bus_oneline}
\end{figure}

\begin{figure}
    \centering
    \includegraphics[width=1\linewidth]{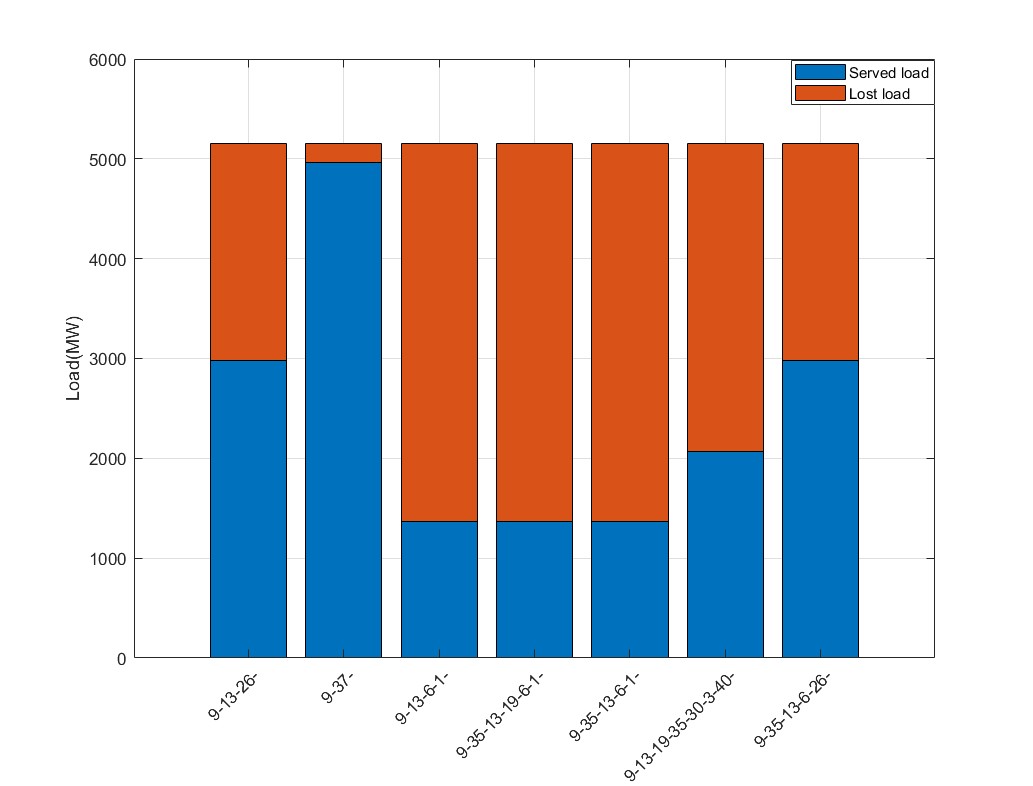}
    \caption{Served load percentage for selected diverse cascading routes.}
    \label{fig:route_diverse}
\end{figure}

\begin{figure}
    \centering
    \includegraphics[width=1\linewidth]{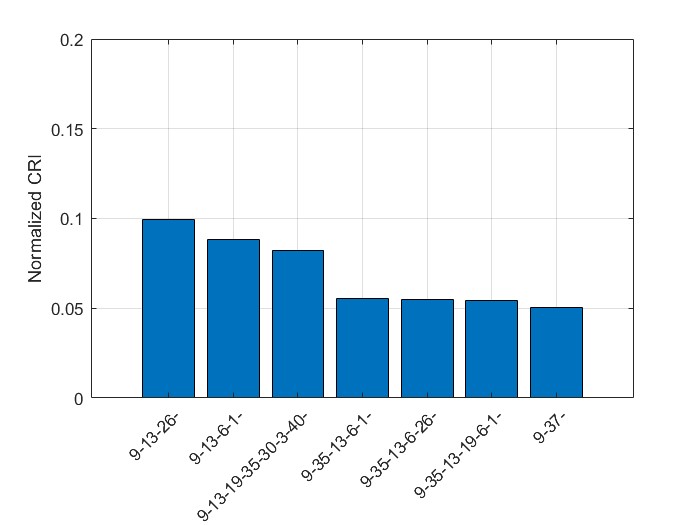}
    \caption{Normalized CRI for selected diverse cascading routes.}
    \label{fig:cri_routes}
\end{figure}

\begin{figure}
    \centering
    \includegraphics[width=1\linewidth]{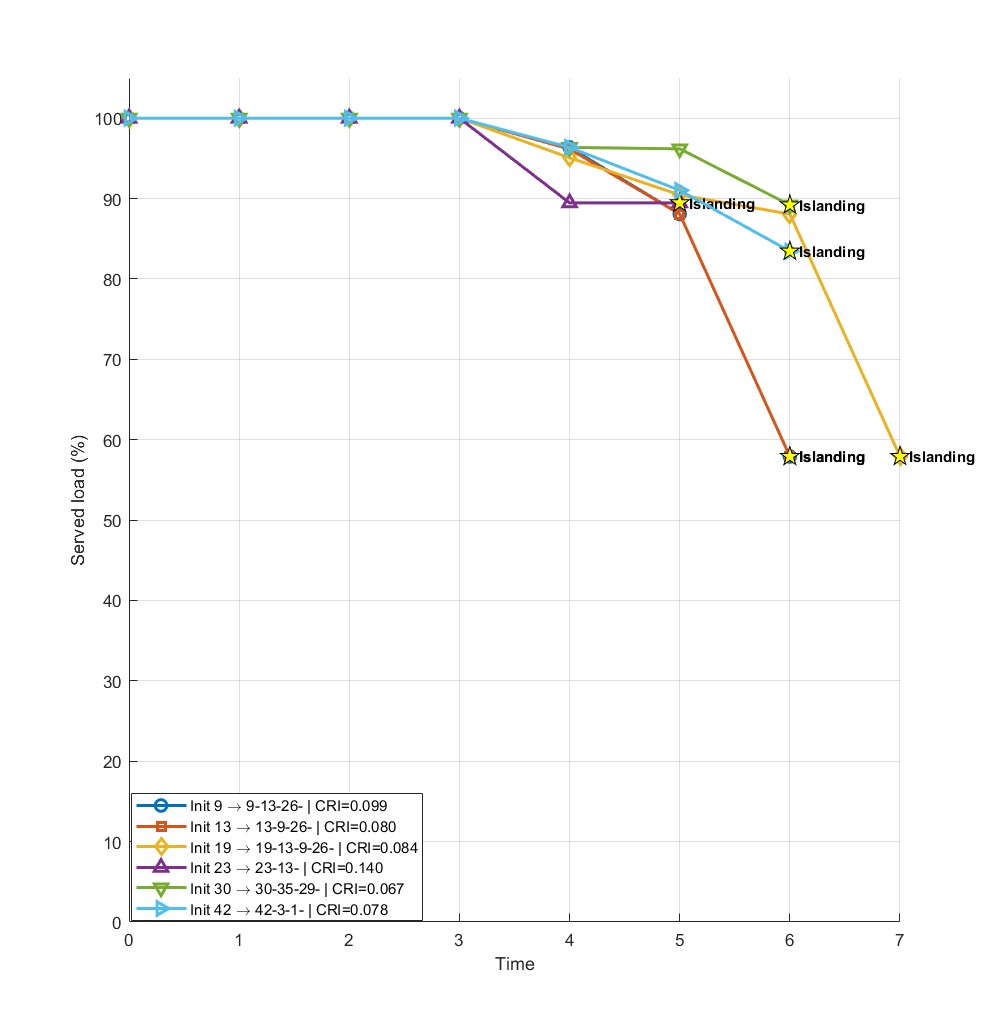}
    \caption{Served load percentage for the highest-CRI route associated with each initiating outage.}
    \label{fig:best_routes_multi}
\end{figure}

%The proposed cascading contingency analysis framework is evaluated on the IEEE 39-bus test system in Fig.~\ref{fig:39bus_oneline}. The IEEE 39-bus system consists of 46 transmission lines and its  detailed electrical network is provided in \cite{pai89}. 
%
The proposed cascading contingency analysis framework is validated using the IEEE 39-bus test system, shown in Fig.~\ref{fig:39bus_oneline}. The test system comprises 39 buses, 10 generators, and 46 transmission lines, and serves as a widely adopted benchmark for power system stability and reliability studies. Detailed electrical and operational data for the network can be found in \cite{pai89}.
Bus 39 is selected as the slack. Post-contingency operating conditions are evaluated using the DC power flow model described in Section~\ref{Sec2}. %which captures active power redistribution, line loading, and load-serving constraints under each contingency mode.

The maximum slack generation capacity is set to $P_{\text{slack}}^{\max}=1200$ MW. A line is considered overloaded when its loading ratio satisfies $|L_\ell|/\bar{L}_\ell>1$. The propagation sensitivity parameter in the logistic outage model is set to $\Delta L=0.05$, which controls the sharpness of the transition from normal operation to failure. To account for stress-dependent degradation, the safe and critical loading thresholds are set to $S_{\text{safe}}=0.95$ and $S_{\text{crit}}=1.50$, respectively, with a degradation factor $\beta=0.15$. A blackout condition is identified when the served load fraction falls below $0.05$. Cascade evolution is terminated when no additional overloaded lines are present or when islanding occurs.

To improve computational efficiency, LODF matrices are computed from the base DC network model. LODF-based screening is used to estimate the impact of the initiating outage and identify a reduced set of candidate lines that may experience stress redistribution. For subsequent cascade stages, DC power flow is recomputed under the updated outage set to accurately evaluate line loadings, islanding conditions, and load shedding. This hybrid approach combines the efficiency of sensitivity-based screening with the accuracy of full post-contingency evaluation.

Fig.~\ref{fig:route_diverse} illustrates the evolution of served load for a set of selected cascading routes corresponding to the initiating outage at bus 9.  The results show that different propagation paths can lead to significantly different system responses. Some routes exhibit gradual degradation in served load, while others experience abrupt reductions due to islanding events. This highlights the importance of modeling cascading failures as multi-stage processes with probabilistic propagation. 

The variation in cascade behavior is further reflected in the corresponding CRI values, as shown in Fig.~\ref{fig:cri_routes}. The CRI is computed using the state-accumulated formulation and normalized by the total system load. Routes with higher CRI values correspond to more severe cascades. In contrast, lower-CRI routes retain a larger portion of the served load throughout the cascade progression. This demonstrates that the CRI effectively captures both the likelihood and severity of cascading failures, and provides a meaningful basis for comparing different propagation scenarios.

To further evaluate system vulnerability, Fig.~\ref{fig:best_routes_multi} compares the served-load evolution associated with the highest-CRI routes for multiple initiating outages (lines 9, 13, 19, 23, 30, and 42). The results show that different initiating outages lead to distinct cascade depths and load-loss patterns. Some initiating events produce relatively moderate degradation, while others trigger more extensive cascading failures with significant load shedding. The marked islanding points indicate stages at which the network separates into disconnected components, leading to additional load loss in accordance with the islanding rules defined in Section~\ref{Sec2}.

The simulation results demonstrate that cascading failure risk depends strongly on both the initiating outage and the subsequent propagation pathway. The hybrid use of LODF-based screening and DC power flow evaluation enables efficient identification of critical transitions, while preserving accuracy in stress and load-shedding estimation. The proposed MC framework captures the probabilistic nature of cascading propagation, and CRI provides a quantitative measure of system vulnerability. 
%by integrating load loss across all stages of the cascade. 
The results also show that the SHS framework can identify high-risk contingencies, distinguish between competing cascade scenarios, and provide insight into the structural and operational factors that drive cascading failures.

\section{Conclusion}
\label{Sec5}

This paper presented an SHS-based framework for cascading contingency analysis in power systems based on a MC representation combined with DC power flow evaluation. The approach models cascading propagation as a sequence of discrete contingency modes, where transition probabilities are derived from line loading conditions and stress-dependent propagation mechanisms. The CRI was introduced to quantify cascading risk based on the likelihood of contingency states and the associated load shedding across cascade evolution.
The proposed framework incorporates a hybrid evaluation strategy in which LODF-based screening is used to efficiently identify candidate outage interactions following an initiating contingency, while DC power flow is employed to accurately compute post-contingency loading, islanding behavior, and load shedding. 
This structure enables scalable exploration of cascading pathways while maintaining physical interpretability of propagation mechanisms.

Simulation results on the IEEE 39-bus system demonstrate that cascading behavior is highly dependent on both the initiating outage and the subsequent propagation path. 
%The results show significant variability in cascade depth, load-loss trajectories, and severity across different routes. 
The CRI metric effectively distinguishes between these scenarios by capturing both intermediate and final-stage impacts. 
%providing a meaningful basis for ranking cascade severity and identifying high-risk contingencies. 
Compared to traditional $N$--$k$ screening, the proposed framework captures the multi-stage and probabilistic nature of cascading failures and reveals critical propagation pathways that may otherwise remain undetected.

Future work will focus on extending the framework beyond the current DC power flow approximation to incorporate more detailed system representations, including AC power flow and voltage-related phenomena. Additional directions include improving the modeling of dependency structures beyond the Markov assumption, integrating data-driven approaches for estimating propagation parameters, and enhancing scalability for large-scale interconnection systems. The incorporation of real-time measurement data and online learning techniques may further enable adaptive risk assessment and support operational decision-making for grid resilience.

\bibliographystyle{ieeetr}
\bibliography{references}

\end{document}